\documentstyle[preprint,aps]{revtex}
\def\gsim {\mbox{\hbox{ \lower-.6ex\hbox{$>$}
\kern-1.1em \lower.5ex\hbox{$\sim$}\kern+.35em}}}
\def\lsim {\mbox{\hbox{ \lower-.6ex\hbox{$<$}
\kern-1.1em \lower.5ex\hbox{$\sim$}\kern+.35em}}}

\begin{document}

\draft

\title{Gravity-Induced Shape Transformations of Vesicles}
\author{Martin Kraus, Udo Seifert and Reinhard Lipowsky
%%%%%%%%%%%%%%%%%%%%%%%%%%%%%%%%%%%%%%%%%%%
%Uncomment following line for LaTeX-article-style
%\\ \\  
%Uncomment following line for RevTeX
} \address{
%%%%%%%%%%%%%%%%%%%%%%%%%%%%%%%%%%%%%%%%%%%
Max-Planck-Institut f\"ur Kolloid- und Grenzfl\"achenforschung\\
Kantstr. 55, 14513 Teltow-Seehof, Germany\\and\\
Institut f\"ur Festk\"orperforschung, Forschungszentrum J\"ulich, \\ 
52425 J\"ulich, Germany\\ 
}

\date{September 13, 1995}

\maketitle
\begin{abstract} 
  We theoretically study the behavior of vesicles filled with a liquid
  of higher density than the surrounding medium, a technique
  frequently used in experiments. In the presence of gravity, these
  vesicles sink to the bottom of the container, and eventually adhere
  even on non - attractive substrates.  The strong size-dependence of
  the gravitational energy makes large parts of the phase diagram
  accessible to experiments even for small density differences. For
  relatively large volume, non-axisymmetric bound shapes are
  explicitely calculated and shown to be stable. Osmotic deflation of
  such a vesicle leads back to axisymmetric shapes, and, finally, to a
  collapsed state of the vesicle.
\end{abstract}

\vskip 1cm
\pacs{PACS numbers: 68.10.-m, 82.70.-y, 87.22.Bt}

%%%%%%%%%%%%%%%%%%%%%%%%%%%%%%%%%%%%%%%%%%%
%   Introduction                          %
%%%%%%%%%%%%%%%%%%%%%%%%%%%%%%%%%%%%%%%%%%%

Lipid  vesicles are simple models for many membrane-bounded
compartments occuring in biology, such as cells or transport vesicles
\cite{lipo94c}. From a physical point of view, they can be understood
as flexible closed surfaces whose shapes and dynamics are controlled
primarily by the  bending energy. Frequently, however, additional
interactions are relevant for a particular experimental situation.
The interactions of two vesicles have been studied with micropipet
experiments \cite{evan90a}. Weak adhesion of a single vesicle to a
substrate can be used to reduce translational and rotational diffusion
and thus to facilitate data analysis \cite{doeb95a}. In this case, one
prefers a small contact area, in order to reduce effects exerted by
the substrate. In other experiments, however, adhering vesicles with a
large contact area serve as a model of a bound planar membrane whose
fluctuations can be analyzed \cite{raed93a,krau94,seif95d}.

In order to stabilize the vesicle at the bottom of the measurement
chamber, a difference in density between the fluids inside and outside
the vesicle is often employed. Mostly, this is done by solvation of
different sugars with equal osmolarity, but different specific weights
\cite{doeb95a}.  However, while bending and adhesion energies have
been taken into account, gravitational energies have so far been
neglected in the analysis of experiments \cite{raed93a,doeb95a}. We
will show that this is not justified even for the small density
differences usually employed.

A simple scaling argument reveals that the contribution of the
gravitational energy cannot be neglected for large vesicles: Whereas
the curvature energy is scale invariant \cite{helf73}, adhesion
energies behave as $F_{\rm adh} \propto R_0^2$. However, the
gravitational energy scales as $F_{\rm grav} \propto R_0^4$, because
it is proportional to the volume multiplied with the height of the
center of mass of the vesicle above the substrate. Here, $R_0 \equiv
(A/(4 \pi))^{1/2}$ is the radius of a sphere with the same area, which
sets the length scale. The energy scale is set by the bending
rigidity, $\kappa$, of the membrane.  We introduce the dimensionless
gravity parameter
\begin{equation}
g \equiv  \frac{g_0\,\Delta\rho\,R_0^4}{\kappa}\,,
  \label{gquer}
\end{equation}
where $g_0 \simeq 9.81$ ms$^{-2}$ is the acceleration of gravity and
$\Delta \rho$ denotes the density difference between the fluids inside
and outside the vesicle. Typical values for the latter are around 0.01
\dots 0.1 g cm$^{-3}$, i.e.\ a few percent of the density of water.
Giant vesicles can reach a size of $R_0 \simeq (5 - 50) \mu $m.  With
$\kappa \simeq 10^{-19} {\rm J} \simeq 25 k_{\rm B} T$, one obtains
values of $0.5 - 5 \times 10^4$ for $g$. Thus, for giant vesicles,
the gravitational energy can be varied over a large range and may be
of the same order of magnitude as the curvature energy.

%%%%%%%%%%%%%%%%%%%%%%%%%%%%%%%%%%%%%%%%%%%
%   Hamiltonian  & Shape Eqns             %
%%%%%%%%%%%%%%%%%%%%%%%%%%%%%%%%%%%%%%%%%%%
 
Previous work has successfully explained the observed shapes of freely
floating giant vesicles in solution by calculating shapes of minimal
bending energy with given constraints
\cite{helf73,deul76,svet89,seif91a,miao94}.  The membrane-substrate
interaction for vesicle adhesion was modeled by a contact energy $W$
\cite{seif90a}, which is justified on a macroscopic scale since the
range of most interactions is orders of magnitude smaller than typical
vesicle sizes. Our effective energy, 
\begin{equation}
  F = \frac{\kappa}{2}\int dA\, (C_1 + C_2 - C_0)^2 + g \int
  dV\,Z - W A_{\rm adh} - W' A_{\rm self}\,,
  \label{helf_ham} 
\end{equation} 
consists of these energies augmented by gravitational and
membrane self-adhesion energies.  Here, $C_1$ and $C_2$ are
the principal curvatures, and $C_0$ denotes the spontaneous curvature
of the membrane. Furthermore, $Z$ denotes the height of a volume
element above the substrate, and $A_{\rm adh}$ is the contact area of
adhesion at the substrate. The membrane adheres to itself with a
contact area $A_{\rm self}$ and contact energy $W'$.

%%%%%%%%%%%%%%%%%%%%%%%%%%%%%%%%%%%%%%%%%%%
% Adhesion Transition & Contact Curvature %
% incl. Gravitation-Induced Adhesion      %
%%%%%%%%%%%%%%%%%%%%%%%%%%%%%%%%%%%%%%%%%%%

The behavior for very small and very large gravitational energies may
be understood by simple arguments.  Vesicles filled with a fluid that
is only slightly denser than the surrounding fluid, i.e.  for small
$g$, will always touch the bottom of the measurement chamber but will
not necessarily form a finite contact area. We will call this state
'pinned' if the vesicle touches the wall only in a single point
\cite{seif91f}.  A contact area of finite size will be formed, as soon
as the cost in bending energy which is necessary in order to form this
area is balanced {\em globally} by a gain in gravitational energy.  In
the axisymmetric case, $C_1$ will denote the principal curvature along
the contour.  The criterion for the transition from the pinned state
to a bound state then follows from the {\em local} boundary condition
at the contact point $S_1$ of the contour \cite{seif90a} as given by
\begin{equation}
R_0 C_1(S_1) = \sqrt{2 w} \,,
  \label{kont_kruemm}
\end{equation}
where $w \equiv W R_0^2 / \kappa$. For vanishing adhesion energy, i.e.
$w=0$, this condition becomes $C_1(S_1) = 0$.  Then, the continuous
adhesion transition between the pinned state and a bound state with
finite contact area happens, when $g$ reaches a critical value where
the vesicle is deformed into a shape which has vanishing mean
curvature at the contact point. Varying $w$ yields a whole line of
adhesion transitions at $g=g_{\rm adh}(w)$.  Numerically, we find
$g_{\rm adh}(w=0) \simeq 0.45$ for $C_0 = 0$ without volume
constraint. For $g=0$, adhesion happens at $w=2$ induced just by the
contact potential \cite{seif90a}.

The behavior for large $g$ is different from the large-$w$-limit.  In
the latter limit, the bending energy may be neglected, and the
vesicles will attain the shape of maximal contact area for given
constraints, which is a spherical cap with a finite contact angle
determined by the Young-Dupr\'e-equation \cite{seif90a}. In the limit
of high gravitational energy, bending is also irrelevant, but now -
given that the volume is fixed - the shape of least energy is that of
a flat disc, as can be seen from an expansion of the gravitational
energy as a function of the reduced volume, $v \equiv V / (4 R_0^3/3 \pi)$,
for small volumes \cite{fgrav_comp}.

%%%%%%%%%%%%%%%%%%%%%%%%%%%%%%%%%%%%%%%%%%%
% Non-Axisymmetric Prolates               %
%%%%%%%%%%%%%%%%%%%%%%%%%%%%%%%%%%%%%%%%%%%

We now want to discuss the details of the shape transformations for
general $g$, which requires the calculation of shapes and their
energies. In order to keep the number of parameters small, we
introduce a few simplifications.  First, we consider vanishing
spontaneous curvature, i.e. $C_0 = 0$, assuming that the different
compositions of the solutions do not affect the symmetry of the
bilayer.  Second, we will focus on $w = 0$ for the membrane-substrate
adhesion and $w'=0$ for the membrane self-adhesion contact
energies \cite{term_adhesion}.  The case with $w \neq 0$ and $w' \neq
0$ will be discussed in more detail in another
publication\cite{krau96}.  Axisymmetric shapes can be calculated
numerically by solving the Euler-Lagrange equations resulting from Eq.
(\ref{helf_ham}) with appropriate boundary conditions
\cite{seif91a,krau96}. A simple argument, however, shows that
non-axisymmetric shapes are also relevant.

For large reduced volume, i.e.\ for shapes close to a sphere, free
prolate vesicles have smaller energy than the corresponding discocytes
\cite{seif91a}. When these shapes adhere under gravity, they will
orient their long axis parallel to the wall. For any finite $g$, the
vesicle will be flattened, thus giving rise to non-axisymmetric
shapes. In order to conserve volume, the asphericity in the plane
parallel to the wall has to be reduced. As discussed above, we expect
the limit shapes for large density difference to be axisymmetric with
the symmetry axis perpendicular to the wall. In general, such an
axisymmetric shape should be reached asymptotically for large $g$.  In
addition, we expect a discontinuous shape transition between adhering
discocytes and non-axisymmetric prolates for small $g$ and $v \gsim
0.65$, since the free vesicle with $C_0=0$ exhibits a discontinuous
transition between discocytes and prolates at $v \simeq 0.65$
\cite{seif91a}.

In order to calculate the energy of these non-axisymmetric shapes a
numerical method can be employed, which minimizes the discretized
curvature energy of a triangulated surface subject to given
constraints. We have used Brakke's Surface Evolver \cite{evolv_ftp}
program, which also allows incorporation of the gravitational energy
easily. Since numerical minimization for a hard wall constraint is
more problematic than for a soft wall, we have used the latter by
exposing the vesicle to the additional potential $V_{\rm w}(z) =
V_{0\rm w} \exp(-Z / Z_0)$, with $V_{0\rm w} / \kappa =5$ and $Z_0/R_0
= 0.1$ which leads to a numerically stable algorithm and induces only
minor deformation of the vesicle shape.

The Surface Evolver data shown in Fig. \ref{nonax_fig} indicate a
'tricritical' point at $v = v_{\rm tr} \simeq 0.88$. Hysteresis
effects indicating a discontinuous transition are found only at $v <
v_{\rm tr}$.  In the limit of small $g$, the limit of metastability
for the discocytes at $v=0.75$ coincides with the results of an
explicit stability analysis for free vesicles \cite{niko94}. At the
spinodals, the energy difference between the unstable and the stable
shape is of the order of several $k_{\rm B} T$; e.g. at $v=0.7$,
$g=11$, the energy of the prolates is $1.50 \kappa \simeq 35 k_{\rm B}
T$ above the energy of the stable discocytes.

In the limit $v \approx 1$, i.e. close to the sphere, we have also
calculated non-axisymmetric shapes using two simple approximations,
whose results confirm the Evolver data and the existence of a
continuous transition in this regime. In the first approach, we expand
the curvature energy and geometrical quantities such as area, volume
and center of mass in spherical harmonics up to $l=2$ along the lines
of Ref.  \cite{miln87}. The influence of gravity is computed under the
assumption of ellipsoidal shapes, so that this method is only useful
for volumes $v \gsim 0.98$. In the second approach, we consider the
restricted set of shapes generated by ellipsoids deformed by special
conformal transformations \cite{seif91c}. Calculation of both
curvature and gravitational energies is now exact, but smaller-volume
shapes are not very well described by the restricted variation. Both
approximations predict a continuous symmetry-breaking transition at
finite $g$, and give an approximation for the transition line
$g^*(v)$, but fail to produce the tricritical point necessary to
change to a discontinuous transition at smaller volumes.  At $v
\approx 1$, however, both methods predict that $g^*(v)$ goes to zero.
We conclude that for large $v$ and intermediate $g$ there are no
stable adhering prolates.

For the general case of non-zero spontaneous curvature, the phase
diagram becomes more complex \cite{krau96}. The trends, however, can
be understood in analogy to the effect of spontaneous curvature on
free vesicles \cite{seif91a}.  Negative $C_0$ stabilizes bound
discocytes and stomatocytes, while positive $C_0$ stabilizes bound
prolates, and, finally, pears and budded shapes.  If one expands the
shapes for $g \neq 0$ around the shapes at $g=0$, one finds that the
amplitudes of the deformation depend linearly on $g$. The prefactors,
however, are a non-monotonic function of $C_0$.

%%%%%%%%%%%%%%%%%%%%%%%%%%%%%%%%%%%%%%%%%%%
% Collaps of Discocytes and Stoma's       %
%%%%%%%%%%%%%%%%%%%%%%%%%%%%%%%%%%%%%%%%%%%

For small $v$, new phenomena occur: (i) Stomatocytes enter the phase
diagram; (ii) The self-avoidance of the membrane has to be taken into
account.  Free discocytes with $C_0 = 0$ self-intersect at their
symmetry axis for $v=0.515$ \cite{seif91a}, while adhering discocytes
self-intersect for $v$ between $0.4$ and $0.5$, depending on $g$ and
$w$.  The physical shapes for even smaller volumes involve
membrane self-adhesion (shapes, which we denote as 'collapsed' in
Fig. \ref{nonax_fig}) even if there is no explicit membrane
self-adhesion energy.  For vanishing or small self-adhesion energy, the cost in
bending energy for forming a finite self-adhesion area is high at the
contact point, and there is a region of 'self-pinned' shapes in the
phase diagram. If the self-adhesion energy is large, the possibility of a
first-order collapse transition arises.  A shape sequence involving
such a collapse transition as generated by osmotic deflation 
(see below) is shown in Fig.  \ref{fig:shapes_coll}.

For $g=0$ and reduced volume $v \leq 0.45$, adhering stomatocytes are
stable when the adhesion energy, $w$, exceeds a critical threshold
value \cite{seif90a}.  As these shapes rise higher above the substrate
than the discocytes of equal volume, they will not be stable for large
$g$. With increasing $g$, adhering stomatocytes redistribute their
volume towards the substrate. As a consequence, the neck at the top of
the vesicle closes. For all membrane self-adhesion energies
$w' > 0$, we find a discontinuous self-adhesion transition. At the top of
the vesicle, a finite area around the infinitesimal neck then sticks
together.  The detailed order and sequence of the transitions between
collapsed and non-collapsed, free and adhering stomatocytes /
discocytes in the small-volume regime depends crucially on the values
of $w$ and $w'$ and will be discussed elsewhere \cite{krau96}. The
energy diagram and shape sequence of a trajectory with fixed $g$ is
shown in Fig.  \ref{g=2_pd}.  The phase diagram might involve even
more complicated shapes such as self-adhering non-axisymmetric shapes
or shapes that self-adhere in more than one place.  This should be
kept in mind when one tries to understand the conformation of swollen
lipid and vesicle-like structures emerging near a substrate.  Many
conformations formerly ascribed to defects in the membrane might be
stable states of pure lipid membranes involving self-adhesion.

%%%%%%%%%%%%%%%%%%%%%%%%%%%%%%%%%%%%%%%%%%%
% Experimental Realizations               %
%%%%%%%%%%%%%%%%%%%%%%%%%%%%%%%%%%%%%%%%%%%

There are several possibilities for experimentally scanning the phase
diagram with a single vesicle.  Raising the temperature will expand
the membrane much more than the liquid. In this case, the actual
volume and the density difference remain constant, but the reduced
volume, $v$, will decrease, while $g$ will increase because of its
dependence on $R_0$. A trajectory starting at a point $(g,v) = (g_s,
v_s)$ will thus continue as $g(v) = g_s (v_s / v)^{4/3}$.

Alternatively, one may vary the volume by exchanging the sugar
concentration, $X_{ex}$, of the exterior fluid, while keeping the
number of dissolved osmotically active particles inside the vesicle,
$N_{in} = V X_{in}$, constant. The volume will then adjust in such a
way that the osmotic pressure, $\Pi = k_{\rm B} T (X_{ex} - N_{in} /
V)$, vanishes up to a negligible contribution of the order of $\kappa /
R_0^3$.  In this case, an increase in the sugar content of the
exterior fluid will raise (i) the density of the exterior fluid
directly, and (ii) the density of the fluid enclosed in the vesicle by
osmotically reducing its volume.  The change of $g$ with $X_{ex}$ thus
depends on experimental details.  If $m_{in,ex}$ denotes the molecular
masses of the sugars inside and outside the vesicle, one obtains
\begin{equation}
  \label{g_osmo}
  g = \frac{g_0\, R_0^4}{\kappa} 
  \left( \frac{N_{in}}{V} m_{in} - X_{ex} m_{ex} \right)
  \approx \frac{g_0\, R_0^4\, X_{ex}}{\kappa} \left( m_{in} - m_{ex} \right)
\label{g_osm}
\end{equation}
in the limit of small $\Pi$.  Even the sign of the response of the density
difference to the increase in sugar concentration, $\partial g /
\partial X_{ex}$, depends on the types of sugars and the other
osmotically active substances involved. It vanishes for
$m_{in}=m_{ex}$.  A trajectory starting at a point $(g_s, v_s)$ will
continue as $g(v) = g_s (v_s / v)$.  Alternatively, one may vary only
the density difference by {\em exchanging} sugars of different
molecular weights in the exterior fluid, while keeping $X_{ex}$ and
the volume constant.

%%%%%%%%%%%%%%%%%%%%%%%%%%%%%%%%%%%%%%%%%%%
% Discussion                              %
%%%%%%%%%%%%%%%%%%%%%%%%%%%%%%%%%%%%%%%%%%%

In conclusion, we have shown that additional energies are necessary
for an experimentally realistic description of adhering vesicles.
Gravity leads to non-axisymmetric shapes, which show continuous
transitions to axisymmetric large-$g$ shapes.  For small volumes, the
self-avoidance of the membrane and the associated self-adhesion energy
lead to a large variety of 'collapsed' shapes. By osmotic deflation or
exchange of sugars it is possible to study these transitions with a
single vesicle.

Stimulating discussions with F. J\"ulicher and W. Wintz are gratefully
acknowledged. We also thank H.G.  D\"obereiner, W. Fenzl, J. R\"adler
and E. Sackmann for sharing their knowledge on the experimental
aspects of this work with us, and K.Brakke for help with Surface
Evolver.

\begin{figure}
\caption{Phase diagram for adhering vesicles as a function of gravitational
  parameter, $g$, and reduced volume, $v$. Non-axisymmetric shapes and
  the large-volume transition are computed using the Surface Evolver
  program, while axisymmetric shapes are solutions of the Euler -
  Lagrange equations with $w=0$. The transition $g^*(v)$ between the
  non-axisymmetric prolates and the discocytes is discontinuous for
  $v <  v_{\rm tr} \simeq 0.88$ (full line) and continuous for $v
  > v_{\rm tr}$ (dashed) separated by a tricritical point (black
  dot). For the discontinuous transition, the limit curves of
  metastable states or spinodals are shown in the figure by dashed
  lines. At $g \lsim 1$ there is an additional transition from
  'pinned' vesicles to those adhering in a finite area; the transition
  line is not shown. On the small-$v$ side of the phase diagram,
  collapsed shapes become relevant.}
\label{nonax_fig}
\end{figure}

\begin{figure}
\caption{Sequence of shapes for collapse transition of discocytes.
  Shapes are taken from a typical osmotic trajectory generated by
  varying the sugar content of the exterior fluid (Eq.
  (\protect\ref{g_osm})).  Parameters approximately matching a typical
  experimental situation \protect\cite{raed93a} with $R_0 \simeq 5
  \mu$m and concentrations of the order 100 mM sucrose or glucose are:
  $w = w' = 0$, $N_in=10^{11}$, $g_0 R_0 m_{in}/\kappa=10^{-10}$,
  $m_{ex} = 0.5 m_{in}$, $x_o R_0^3 \simeq 4.3 \cdot 10^{-10} (1),4.8 \cdot
  10^{-10}$ (2), and $6.3 \cdot 10^{-10}$ (3), leading to volumes
  $v = 0.555$ (1), $v=0.500$ (2), and $v=0.379$ (3), respectively.}
\label{fig:shapes_coll}
\end{figure}

\begin{figure}
\caption{Energy $F$  
  for adhering vesicles at $g=2$, $w=0$, and $w'=0.05$ as a function
  of the reduced volume, $v$. Note that lines of constant $F$ are
  shown diagonally and are not orthogonal to lines of constant
  $v$. Stable shapes for the various volumes and their transition lines
  (dashed) are shown in the lower part. All shapes adhere to the
  substrate. The non-axisymmetric shapes have been calculated only at
  the marked points.}
\label{g=2_pd} 
\end{figure}


\begin{thebibliography}{10}
\bibitem{fgrav_comp}
The shape of a vesicle for large $g$ can be approximated by a 'coin'-shaped
  flat disc of fixed area $A=4 \pi R_0^2$. Denoting the gravitational part of
  its free energy by $F_{\rm grav}$, we obtain $F_{\rm grav}/{\kappa g} = 4/9
  \pi v^2 + \sqrt{32}/27 \pi v^3 + 4/27 \pi v^4 + \dots$. This is lower than
  the result $F_{\rm grav}/{\kappa g} = 32/27 \pi v^2 + 64/729 \pi v^4 + \dots$
  for the analogous spherical cap.
\bibitem{term_adhesion}
Strictly speaking, the vesicle has non-adhesive contact with the substrate in
  the case $w = 0$. Keeping the generalization to $w \neq 0$ in mind, we will
  also call this case 'adhesion'.
\end{thebibliography}
\end{document}